\documentclass{webofc}
\usepackage[varg]{txfonts}   
\begin{document}
\title{Strange Baryons in Nuclei and Neutron Stars}
%
%

\author{\firstname{Laura} \lastname{Tolos}\inst{1,2,3,4}\fnsep\thanks{\email{
tolos@ice.csic.es}}}

\institute{Institute of Space Sciences, Campus UAB,  Carrer de Can Magrans, 08193 Barcelona, Spain
\and
Institut d'Estudis Espacials de Catalunya, 08034 Barcelona, Spain
\and
Faculty  of  Science  and  Technology,  University  of  Stavanger,  4036  Stavanger,  Norway
\and
Frankfurt Institute for Advanced Studies, Ruth-Moufang-Str. 1, 60438 Frankfurt am Main, Germany
          }

\abstract{%
In this contribution we review the dynamics of hyperons with nucleons and nuclear matter, paying a special attention to hypernuclei. We also discuss the presence of hyperons in the inner core of neutron stars and the consequences for the structure of these compact stars.
}
\maketitle
\section{Introduction}
\label{intro}
The dynamics of hyperons with nucleons and nuclear matter has been object of high interest in connection with the formation of hypernuclei as well as the probable existence of hyperons in neutron stars \cite{Tolos:2020aln}.
In order to understand this dynamics, first we need to address the features of the hyperon-nucleon (YN) and hyperon-hyperon (YY) interactions, so as to comprehend how few-body systems such as hypernuclei are formed and how hyperons interact with other nucleons and hyperons, in particular in the core of neutron stars.

In this contribution we briefly review the status of YN and YY interactions, paying a special attention to hypernuclei. We then address hyperons in dense nuclear matter within the chiral effective theory ($\chi$EFT) framework. The final goal is to discuss the probable presence of hyperons in the core of neutron stars and the consequences for the structure of these compact objects. The present contribution represents a brief summary of part of the  review of Ref.~\cite{Tolos:2020aln}.

\section{YN and YY interactions from a theoretical perspective}
\label{theory}

Over the years a lot of effort has been made from the theoretical point of view to describe the YN and YY interactions. The theoretical approaches can be grouped in meson-exchange models, $\chi$EFT approaches, 
computations on the lattice (LQCD), low-momentum schemes and quark-model potentials.

The interaction between two baryons in {\it meson-exchange models} is mediated by the exchange of mesons. Starting from the nucleon-nucleon (NN) meson-exchange model, SU(3)$_{\rm flavor}$ symmetry is assumed
to determine the YN and YY interactions. Those schemes include the J\"ulich \cite{Haidenbauer:2005zh} and Nijmegen potentials \cite{Rijken:2010zzb}.

Within chiral effective field theory, the J\"ulich-Bonn-Munich group has built the YN and YY interactions starting from their previous $\chi$EFT approach for the NN interaction \cite{Polinder:2006zh,Haidenbauer:2013oca,Haidenbauer:2019boi}.

As for LQCD, Monte Carlo techniques are used to solve the QCD path integral over the quark and gluon fields at each point of a four-dimensional space-time grid. This effort has been lead by the HALQCD \cite{halqcd} and the NPLQCD \cite{nplqcd} collaborations.

And, finally, other schemes for the description of the YN and YY interactions include {\it low-momentum interactions}  and {\it quark-model potentials}. Whereas the former aims at determining a universal effective low-momentum potential for YN and YY by means of renormalization-group methods \cite{Schaefer:2005fi},  the latter constructs the YN and YY interactions within constituent quark models \cite{Fujiwara:2006yh}.

\section{Hypernuclei}
\label{hypernuclei}

Hypernuclei are bound systems composed of neutron, protons, and one or more hyperons. Over the years their properties have been addressed experimentally and theoretically, so as to relate them with the underlying YN and YY interactions \cite{Gal:2016boi}. 

In particular, in the last years the determination of the hypertriton ($^3_{\Lambda}$H) lifetime has received a lot of attention. The separation energy of the $\Lambda$ in this hypernucleus is about $130$ keV, and this results in an RMS radius of $10.6$ fm for the hypertriton. A very low binding energy suggests a small change of the $\Lambda$ wave function in a nucleus and, hence, the lifetime of the hypertriton should be very close to that of the free $\Lambda$ hyperon. However, several measurements reported a shorter lifetime for $^3_{\Lambda}$H and this gave rise to the so called {\it hypertriton lifetime puzzle}.  

The recent measurement of the STAR collaboration \cite{STAR:2019wjm} suggests a value of the binding energy 
about three times larger than the original benchmark \cite{Juric:1973zq}. However, measurements of the hypertriton lifetime by ALICE \cite{ALICE:2019vlx} give a small value of the hypertriton binding energy, indicating consistency with the free $\Lambda$ lifetime.

\section{Hyperons in dense matter within $\chi$EFT framework}
\label{dense}

During the past decade the YN interaction has been extensively studied within the $\chi$EFT. In this framework the YN interaction has been obtained within the chiral expansion, improving calculations systematically by moving to higher orders in the Weinberg power counting \cite{Polinder:2006zh,Haidenbauer:2013oca,Haidenbauer:2019boi}. 

Using, as kernel, the  LO and NLO contributions for YN,  the properties of $\Lambda$ and $\Sigma$ in dense matter have been obtained within the Brueckner-Hartree-Fock framework~\cite{Haidenbauer:2014uua, Petschauer:2015nea}. Whereas 
the $\Sigma$-nuclear potential is found to be repulsive, the $\Lambda$ single-particle potential is in good qualitative agreement with the empirical values extracted from hypernuclear data, becoming repulsive about two to three-times saturation density.  The inclusion of the three-body forces for the $\Lambda$-nuclear interaction in dense matter \cite{Gerstung:2020ktv} gives an extra repulsion at large densities. As we will discuss in the next section, this might help to solve the so-called {\it hyperon puzzle}.

\section{Hyperons inside Neutron Stars}
\label{neutronstars}

Neutron stars are one of the most compact known stellar objects and, therefore, turn out to be a unique laboratory for testing dense matter \cite{Watts:2016uzu,Watts:2018iom}. The mass and radius of these stellar objects depend on the properties of matter in their interior and, thus, are strongly correlated to the equation of state (EoS) of the core. Very accurate observations of $2M_{\odot}$ neutron stars have been determined \cite{Demorest:2010bx,Antoniadis:2013pzd,2020NatAs...4...72C}, whereas 
both mass and radius for certain neutron stars have been obtained very recently by NICER (Neutron star Interior Composition ExploreR) \cite{Miller:2019cac,Riley:2019yda,Riley:2021pdl,Miller:2021qha}. 

The composition of the core of neutron stars is determined imposing equilibrium against weak interaction processes, the so-called $\beta$-stability. Whereas the core of neutron stars has been traditionally modelled by a uniform fluid of neutron rich matter in  $\beta$-equilibrium, other degrees of freedom could be expected, such as hyperons. This is due to the fact that, on the one hand, the value of the nucleon chemical potential increases rapidly with density and, on the other hand, the density in the core of neutron stars is very high. 

The probable appearance of hyperons in the core of neutron stars would affect the structure of these compact objects. Indeed, the EoS and, hence, the pressure becomes softer with respect to the case when only nucleons are present. The reason behind is that the addition of one specie opens a set of new available low-energy states that can be filled, hence lowering the total energy (and pressure) of the system. As the total pressure of the star is lowered, the gravitational pull is reduced in order to maintain the hydrostatic equilibrium. As a consequence, the less pressure, the less mass a neutron star needs to sustain, leading to masses for neutron stars below the $2M_{\odot}$ observations. This is usually referred as {\it the hyperon puzzle}. 
 
Several solutions have been put forward  so as to obtain  $2M_{\odot}$ neutron stars with hyperons in the core. Those solutions move from stiffer YN and YY interactions, stiffening induced by hyperonic three-body forces, the appearance of 
$\Delta$ baryons, meson condensates or a phase transition to quark matter below the hyperon onset to the existence of dark matter in the core of neutron stars or the need of modified gravity theories for the description of neutron stars (see \cite{Vidana:2018bdi,Tolos:2020aln} and references therein). 

\section{Summary}

We have briefly reviewed the dynamics of hyperons with nucleons and nuclear matter, paying a special attention to hypernuclei and, in particular, to the hypertriton lifetime puzzle. We have also discussed the hyperon puzzle in the context of neutron stars, as hyperons would appear in the core of neutron stars, softening the equation of state and leading to neutron stars with masses below the $2M_{\odot}$  observations.

\section*{Acknowledgements}
The author wishes to acknowledge support from the Spanish Ministerio de Ciencia e Innovaci\'on and the European Regional
Development Fund (ERDF) under contract PID2019-110165GB-I00 (Ref.10.13039/501100011033); the Consejo Superior de Investigaciones Cient\'ificas under Project Ref.~202050I008; the Deutsche Forschungsgemeinschaft (DFG, German research Foundation) under Project Nr. 411563442 and Project Nr. 315477589 - TRR 211 (Strong-interaction matter under extreme conditions); the EU STRONG-2020 project under the program H2020-INFRAIA- 2018-1, grant agreement no. 824093; and the Generalitat Valenciana under contract PROMETEO/2020/023.

%
 \bibliography{sqm2021_tolos_biblio}
%
%
%
%

\end{document}